\documentclass[[aps,pra,pdf,superscriptaddress,amsmath,amssymb,amsfonts,twocolumn,showpacs,nofootinbib, longbibliography]{revtex4-1}

\usepackage{amssymb}
\usepackage{eqnarray,amsmath}
\newcommand{\abs}[1]{\left| #1 \right|} 
\usepackage{graphicx}
\usepackage{epsfig}
\usepackage{dcolumn}
\usepackage{bm}
\usepackage{braket}
\usepackage{amsmath}
\usepackage{physics}
\usepackage{csquotes}
\usepackage{mathtools}
\usepackage{graphicx,color,xcolor,colortbl}

\usepackage[colorlinks=true,
linkcolor=blue,
filecolor=blue,      
urlcolor=blue,
citecolor=blue]{hyperref}

\usepackage[normalem]{ulem}

\definecolor{Gray}{gray}{0.85}
\definecolor{LightCyan}{rgb}{0.88,1,1}

\newcolumntype{a}{>{\columncolor{Gray}}c}
\newcolumntype{b}{>{\columncolor{white}}c}

\begin{document}
\author{M. Schubert}
\email{malte.schubert@fysik.lu.se}
\affiliation{Mathematical Physics and NanoLund, Lund University, Box 118, 22100 Lund, Sweden}
\author{K. Mukherjee}
\email{koushik.mukherjee@fysik.lu.se}
\affiliation{Mathematical Physics and NanoLund, Lund University, Box 118, 22100 Lund, Sweden}
\author{T. Pfau}
\affiliation{Physikalisches Institut and Center for Integrated Quantum Science and Technology, 
Universität Stuttgart, Pfaffenwaldring 57, 70569 Stuttgart, Germany}

\author{S. M. Reimann}
\affiliation{Mathematical Physics and NanoLund, Lund University, Box 118, 22100 Lund, Sweden}

\title{Josephson vortices and persistent current in a double-ring supersolid system}
\begin{abstract}

We theoretically investigate the  properties of  ultra-cold dipolar atoms in radially coupled, concentric annular traps created by a potential barrier. The non-rotating ground-state phases are investigated across the superfluid-supersolid phase transition, revealing a particle imbalance between the two rings and a preferential density modulation in the outer ring in the absence of rotation. Near the phase transition on the superfluid side, applying rotation can induce density modulations in either ring, depending on the angular momentum and barrier strength. For low angular momentum, such rotation-induced density modulation forms in the outer ring, while for high angular momentum and weak barriers, it emerges in the inner ring. Rotation can lead to persistent currents and the nucleation of a vortex residing either at the center (central vortex) or at the ring junction (Josephson vortex). Josephson vortices can also form at the junctions of the localized density sites induced by rotation in the inner ring, a behavior that is unique to our system. By switching off the trap and allowing the system to expand, distinct interference patterns emerge, which can be analyzed to identify and distinguish between various vortex configurations, and thus can be observed in current state-of-the-art experiments.

\end{abstract}
\maketitle

\section{Introduction}
Superfluidity is well-known to be closely related to the phenomenon of Bose-Einstein condensation (BEC)~\cite{Gross1957, Bloch1973, Leggett1999_SF} and manifests through the existence of vortices and persistent currents (see the review~\cite{Fetter_2009}). 
In analogy to superconducting rings~\cite{Bloch1965, Mateev2002}, multiply-connected atomic condensates in 
toroidal traps may exhibit metastable flow~\cite{Ryu2007,Ramanathan2011_prl,Moulder2012,Wright2013_prl,Beattie2013,Murray2013,Eckel2014,Jendrzejewski2014_prl,YGuo2020} (see also the recent review~\cite{Polo2024}). 
Likewise, the Josephson effect (originally discovered in superconducting systems~\cite{Josephson1962}) may govern the tunneling between   purely superfluid (SF) states that are weakly linked by a junction formed by an external potential.  For singly-connected systems, the atomic analogue of the Josephson effect has been intensively studied, see, e.g., Refs.~\cite{Smerzi_1997, Marino_1999, Albiez2005, Gati_2007, Levy2007, LeBlanc_2011, Gallemi_2016a, Gallemi_2016b,Spagnolli2017, Pigneur_2018}. 
Particularly interesting however is the combination of the Josephson effect and persistent flow that can be achieved by trapping a BEC in a double (or multiple) ring geometry, arranged coaxially or coplanarly ~\cite{Lesanovsky2007,Brand_JV_2009,Brand2010, Malet_2010, Aghamalyan_2013, Zhang_2013, su_2013, Polo_2016, Bland_2022_double_ring, Bazhan2022_JV, Borysenko2024, Chaika2024, Mukherjee2025Selective}. The coupling between the rings  across an azimuthally symmetric barrier may then lead to an intriguing interplay between Josephson tunneling and persistent currents (PCs) in the system. 
A distinctive feature of such multi-ring potentials is their ability to support either identical or distinctly quantized flows across the junctions, with some or all of the rings carrying quantized angular momentum. The phase difference between the rings leads to the formation of vortices at the Josephson barriers, commonly referred to as Josephson vortices (JVs). They have been observed in superconductors~\cite{Roditchev2015_JV_supercon} and polariton superfluids~\cite{Caputo2019_JV_polariton}, and are also well
studied in BECs of alkali atoms~\cite{Kauriv_JV_2005, Brand_JV_2009, Montgomery_2015, Gallemi_2016b, Oliinyk2020, Bazhan2022_JV, Borysenko2024,Tononi_2024}. 

Dipolar BECs~(as reviewed in~\cite{Lahaye2009, Boettcher2021, Chomaz2022, Mukherjee_review_2023}) add another interesting twist to the physics of JVs and PCs, due to the long-ranged interaction. After first experiments with 
Chromium~\cite{Griesmeier2005,Stuhler2005}, also lanthanides  with larger magnetic dipole moments~\cite{Lu2010, Lu2011, Aikawa_2012, Miyazawa_2022} became of interest, where similarly to a classical Rosensweig transition, regular arrays of droplets may form~\cite{Kadau2016,Ferrier2016}.  Under certain 
conditions, these droplets may phase-coherently overlap  and a periodic solid-like structure may emerge while the  coherent superfluid properties are partly maintained~\cite{Boettcher2019,Tanzi2019a,Chomaz2019, Norcia2021}.
Such  ``supersolid" (SS) state of matter was predicted early on~\cite{Gross1957, *Gross1958, *Yang1962, *Chester1970, *Leggett1970} for helium but remained elusive~\cite{Boninsegni2012}. Unequivocal evidence for its existence  however only came more recently from the above experiments with ultra-cold dysprosium~\cite{Tanzi2019a, Chomaz2019, Boettcher2019, Norcia2021} and erbium~\cite{Chomaz2019}. Subsequent studies analyzed the excitation spectra~\cite{Guo2019, *Natale2019, *Hertkorn2019, *Schmidt2021, *Hertkorn2021supersolidity2D, *Hertkorn2021densityfluctuations, Buchler_ss1_2023,*poli_2d_excitation_2024} associated with the SF-SS transition and provided deeper insights into various dynamical phenomena~\cite{Tanzi2019b, *Ilzhofer2021, *Sohmen2021, *Bland2022_2Dsupersolid, *alana_roton_SS_2023, *Mukherjee_linear_chain, *Mistakidis_tunneling_2024}. Vortices as indicators of superfluidity in the SS state were also  studied~\cite{Roccuzzo2020, Gallemi2020, Sindik2022,Klaus2022, Casotti2024}. 
Interestingly, already long before the realization of SSs, it was suggested that a SF of dipolar atoms polarized perpendicular to the symmetry axis of a toroidal trap will form a self-induced Josephson junction, splitting the SF in two halves on either side of the ring~\cite{Abad_2011b}. More recently, the SS state has been interpreted as an array of such junctions~\cite{Biagioni2024, Platt2024, Beatrice_JE_2025, Alana_JE_2025}, albeit here as a consequence of the SS density modulation. A connection could be drawn between the superfluid fraction as a measure of the density modulation in the SS, and the Josephson effect~\cite{Biagioni2024}. 
For a dipolar SS in a toroidal trap, earlier studied in~\cite{Tengstrand2021,Tengstrand2023,Sindik2024} and yet to be realized experimentally, the absence of inhomogeneities typical of cigar-shaped traps leads to a collective excitation spectrum where first sound, second sound, and Higgs modes can decouple~\cite{Hertkorn_2024}. When persistent current exists, the angular momentum per particle (in the SS being less than unity in units of $\hbar$) is determined by the superfluid fraction~\cite{Tengstrand2021, Tengstrand2023, Sindik2024}. 

\begin{figure}[t!]
\begin{center}
\includegraphics[width=0.45\textwidth]{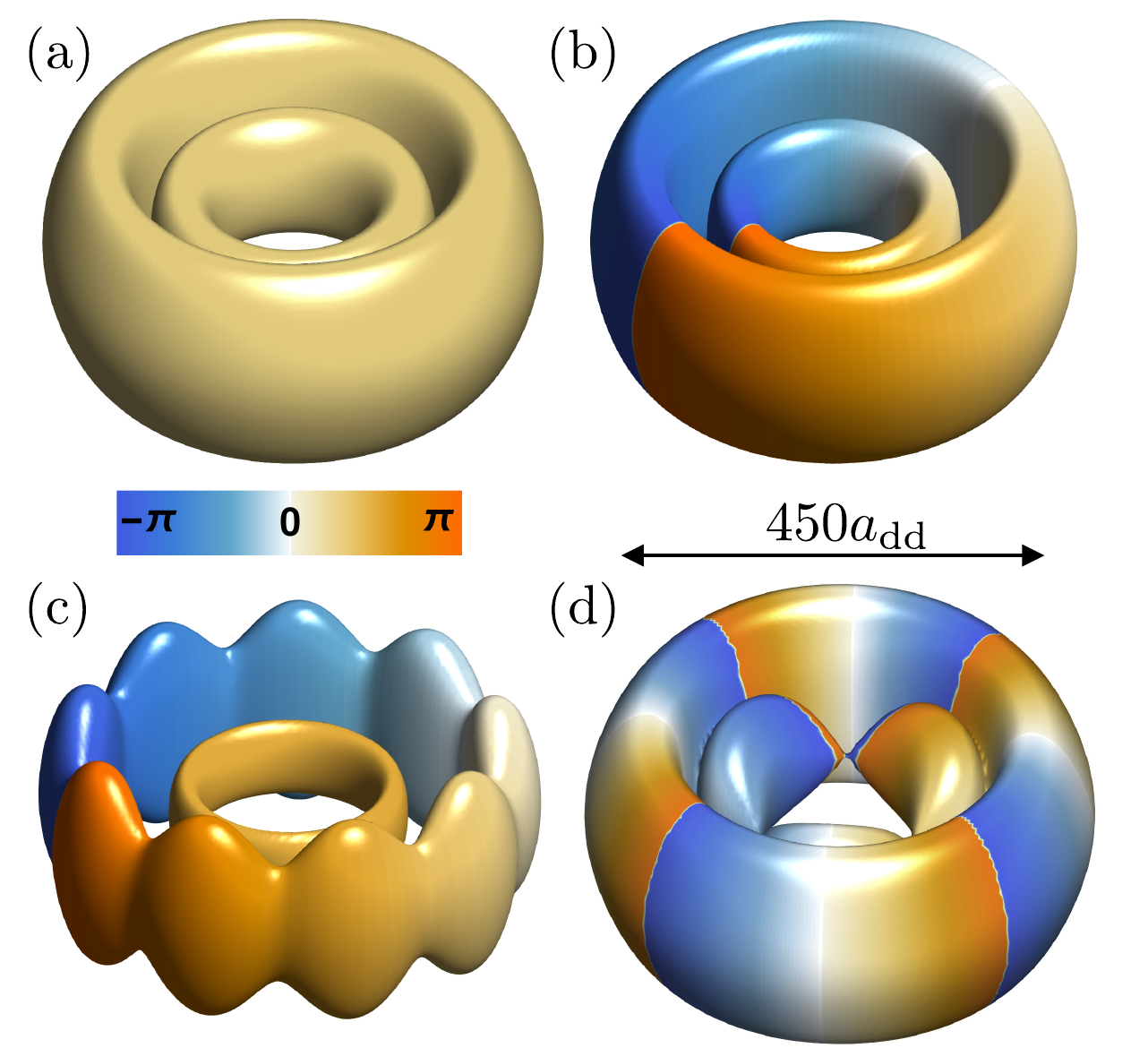}
\end{center}
\caption{Density isosurfaces in ground states showcasing SF states [upper panel, (a), (b)] and SS states [lower panel, (c),(d)]. The color of the isosurface at each position represents the value of the phase associated with the wave function. (a) Non-rotating ground state for $\epsilon_{\rm dd}=1.95$ and $V_{B}=10^4a_{\rm dd}^2$. (b) Rotating ground state with a  CV obtained for the same parameters as in (a), but for a trap rotation with  $\Omega=0.04\omega$. (c) Rotating ground state with a $\rm JV_1$ for  $\epsilon_{\rm dd}=2.05$, $V_{B}=6\times 10^4a_{\rm dd}^2$ and $\Omega=0.04\omega$.  (d) Rotating ground state with three $\text{JV}_2$s. The parameters here are the same as in (a),  but with $\Omega=0.12\omega$. 
The iso-surfaces are plotted at values of $10\%$ of the maximum density for (a)-(c) and at $38\%$ for (d).}
\label{fig_density}
\end{figure}

In this work, motivated by the advantages of the toroidal confinement and the existence of novel phases in dipolar BECs, we investigate dipolar BECs confined in coplanar double rings sharing a common center (as displayed by the density iso-surfaces in Fig.~\ref{fig_density}). 
Pertinent questions are how the spontaneous density modulation of the SS state evolves in a double-ring as the relative dipolar interaction strength increases, how the persistent current develops, and how topological defects emerge in the presence of rotation. 
The system remains in a pure SF state when both rings exhibit azimuthally uniform density profiles, as shown in  Figs.~\ref{fig_density}(a)-(b).  
If a density modulation occurs in one or both rings, 
The system may also form a SS state [Figs.~\ref{fig_density}(c)-(d)], with a density modulation in one or both of the rings. We note that such a SS system contains two types of junctions: (1) one between the rings, produced artificially by the azimuthally symmetric barrier between the rings, and (2) those formed by the atoms that take part in the superfluid flow of the SS. 

The long-range dipolar interaction creates a population difference between the rings, and the supersolid density modulation preferentially appears in the outer ring in the absence of rotation. Near the SF-SS phase transition, a system that initially is in a SF state can form a density modulation when forced to acquire angular momentum through rotation. This may happen either in the outer or inner ring, depending on the system's angular momentum and the barrier strength. For relatively small angular momentum, it is the outer ring that contributes to the formation of the SS, regardless of the barrier strength. For large angular momentum, if the barrier is weak, rotation can facilitate formation of a SS in the inner ring.  

When a critical rotation frequency is exceeded, topological defects such as vortices are nucleated as a consequence of the superfluid properties of the system, and also persistent currents may occur. 
A non-zero density along the azimuthal barrier, i.e. in between the two rings, significantly affects the pathway of vortex nucleation.  We in the following refer to vortices that are located at the junction barrier between the rings as $\rm JV_1 $ [Fig.~\ref{fig_density}(c)]. In this case, metastable persistent current exists only in the outer ring, as we observe only for the isolated rings.  The entire system can exhibit a persistent current when a vortex is located at the center, referred to as a central vortex (CV), which occurs when a density bridge exists between the rings due to a weaker barrier [Fig.~\ref{fig_density}(b)]. In this latter configuration, sufficiently high rotation can induce formation periodic density modulation in the inner ring, with vortices located at the junctions between the density sites [Fig.~\ref{fig_density}(d)]. These JVs are unique to the double-ring dipolar system, where the system’s tendency to form spontaneous density modulation makes their existence possible. We refer to them as $\rm JV_2$. Notably, these structures can be observed in experiments through the interference of different parts of the condensate, producing distinct patterns when the trap is switched off.

The remainder of this paper is organized as follows: We introduce our setup and theoretical framework in Sec.~\ref{model}. Our results are discussed in Sec.~\ref{results}. Specifically, we first examine the static ground-state structures that develop in the double-ring system in Sec.~\ref{non_rot_states}. The rotational dynamics of the system are then analyzed in Sec.~\ref{rot_states} under two distinct configurations: one where the rings remain separated [Sec.~\ref{sep_rings}] and another where they are connected [Sec.~\ref{con_rings}]. A phase diagram explaining the existence of JVs and CVs is presented in Sec.~\ref{phase_dia}. We discuss how interferometric techniques can be used to distinguish between these states in Sec.~\ref{vor_det}. After conclusive remarks and an outlook in Sec.~\ref{conclu}, Appendices~\ref{comdetails} and ~\ref{Derivation} provide some additional details of the numerical simulations and analytical expressions, respectively,  performed in this work.


\section{Model and methods}\label{model}

The confinement setup can be realized using a toroidal potential of radius $r_{0}$, supplemented by a Gaussian potential centered at $r = r_{0}$ (where $r = \sqrt{x^2 + y^2}$) forming an azimuthally symmetric barrier that makes it possible to split the confinement into an inner and an outer ring: 
\begin{equation}
V(\mathbf{r})=\frac{1}{2}M\omega^2\left[ \left(r-r_0\right)^2+\lambda^2 z^2+V_{ B} e^{ -( \frac{r-r_0}{\sigma } )^2 } \right].    
\end{equation}
The potential has two minima located at $r_1 = r_{0} - \sqrt(\sigma^2 \ln{V_{ B}/\sigma^2})$, and $r_2 =
r_{0} + \sqrt(\sigma^2 \ln{V_{ B}/\sigma^2})$. The confinement frequency in the radial plane is given by $\omega/(2\pi)$, while that along $z$ is $\omega_z = \lambda \omega$. The width and strength of the barriers are characterized by the parameters $\sigma$ and $V_{ B}$, respectively. 
In the following, we analyze the rotational properties of dysprosium atoms confined by the above potential. The behavior can be modeled using the usual extended Gross-Pitaevskii equation (eGPE)  $i\hbar\partial\psi/\partial t= \delta E[\psi ]/\delta \psi^*$ with the corresponding energy functional in the non-rotating frame given by
 \begin{eqnarray}\label{GPEEnergy}   
\notag & E=\int dV\bigg( \frac{\hbar^2}{2M}| \nabla \psi|^2  +V|\psi|^2+\frac{1}{2}g|\psi|^4\\ &+\frac{1}{2}g_{\rm dd}|\psi|^2\left(\frac{1-3\cos^2\theta}{|\mathbf{r}|^3}*|\psi|^2\right)+\frac{2}{5}\gamma|\psi|^5  \bigg). 
 \end{eqnarray}
In the rotating frame, the above equation reads \( E(\Omega ) = E - \Omega L \), where \( \Omega \) is the rotation frequency and \( L = \int dV\, \psi^* \hat{L}_z \psi \) is the angular momentum, with \( \hat{L}_z = -i\hbar(x \partial_y - y \partial_x) \) being the angular momentum operator. The contact interaction has strength  $g=4\pi\hbar^2 a/M$ and can be tuned by varying the s-wave scattering length $a$. We denote the particle mass by $M$. The angle $\theta$  is defined as the angle between the position vector and the dipole moment, which is assumed to align with the  $z$-direction. The coefficient of the dipole-dipole interaction (DDI) is $g_{\rm dd} = 4\pi\hbar^2a_{\rm dd}/M$, where $a_{\rm dd}= \mu_{0}\mu^2_{m}M/12\pi \hbar^2$ represents the dipolar length. 
The last term in Eq.~\eqref{GPEEnergy} is the so-called Lee-Huang-Yang (LHY) correction
 where $\gamma = \frac{32}{3}g \sqrt{a^3/\pi} \left(1+\frac{3}{2}\epsilon_{\rm dd}^2\right)$ and $\epsilon_{\rm dd} = a_{\rm dd}/a$~\cite{Lima2011,Lima2012}.
  The ground states are determined by solving eGPE equation using the split-step Fourier method in imaginary time, while real-time evolution is employed to explore the system’s dynamical behavior.  We here consider $N = 10,000$  $^{164}$Dy atoms with $a_{\rm dd} = 130a_{0}$. The trapping frequencies are $(\omega, \omega_z) = (1000, 1700)\, \rm Hz.$ and the width $\sigma  = 10a_{\rm dd}$.

 By varing $V_{ B}$, $\epsilon_{\rm dd}$ and $\Omega$, in the following we systematically explore both the non-rotational and  rotational properties of the system.
When $\Omega$ exceeds a critical value,  JVs and  CVs can emerge in the ground state,  identified by examining the lowest energy $E(L)$ as function of angular momentum $L$ (i.e., the so-called yrast line~\cite{Mottelson1999}). To do this, we shall minimize $E(L) = E + C (L- L_0)^2$, where $C$ is a large number ~\cite{Tengstrand2021}. The energy for a single component in a toroidal setup can be expressed as the sum of a term quadratic in \(L\), arising from the kinetic energy, and another function primarily originating from particle interactions, which is symmetric and periodic in \(L\) when central vortices are generated in the system~\cite{Bloch1973}. A minimum in energy at a value \(L = L_0\) indicates  a metastable persistent current with angular momentum \(L_0\) in the ground state~\cite{Bloch1973,Voigt_1983, Komineas_2005, karkkiainen_persistent_2007}.
To understand the rotational behavior of the system, we examine both the yrast lines (see, for example, Fig.~\ref{rot_fig_sep}(a), Fig.~\ref{rot_fig_con}(a), and Fig.~\ref{YrastFull}) and the angular momentum as a function of $\Omega$ (see, for example, Fig.~\ref{rot_fig_sep}(b), Fig.~\ref{rot_fig_con}(b), and Fig.~\ref{PhaseDia}).

 \begin{figure}[b!]
\begin{center}
\includegraphics[width=0.5\textwidth]{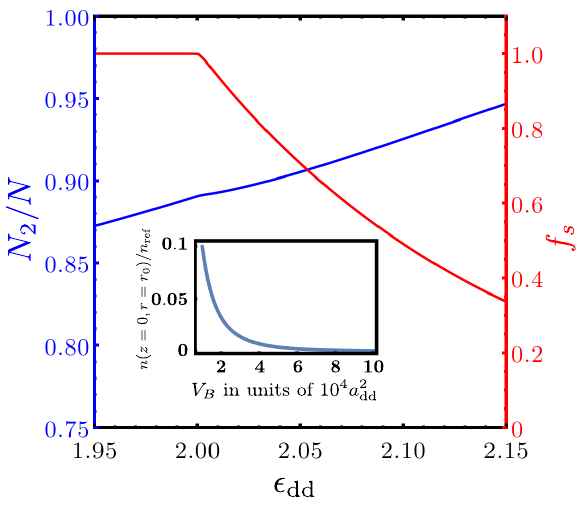}
\end{center}
\caption{The relative population, $N_{2}/N$ [left axis], of the condensate residing in the outer ring and superfluid fraction, $f_s$, [right axis] as a function of the relative dipolar strength $\epsilon_{\rm dd}$ for a barrier strength $V_{B}=10^5a_{\rm dd}^2$. The inset shows the azimuthal symmetric density, $n(r = r_{0}, z = 0)$ for $\epsilon_{\rm dd}=1.95$ as a function of $V_{B}$. The reference density, $n_{\rm ref}$ corresponds to the density at $r = r_{0}$ and $z=0$ when $V_{B} = 0$.}
\label{fig_n2_fs}
\end{figure}

 \section{Results and Discussions}\label{results}

\subsection{Non-rotating ground states}\label{non_rot_states}

Let us first highlight the static ground state properties of the double-ring system, initially focusing  on barrier strength $V_{B} = 10^5a^2_{\rm dd}$, which separates the rings without any density overlap between them. The density configurations are determined by how the dipolar atoms are distributed within the two rings. The particle number of the outer rings, $N_2 = \int_{r>r_0}dV \abs{\psi}^2$, is shown  as a function of $\epsilon_{\rm dd}$ in Fig.~\ref{fig_n2_fs}. It indicates that the outer ring always has a higher population [see also Fig.~\ref{fig_density}] increasing further as the dipolar interaction becomes stronger.  This behavior can be understood by considering both the long-range  nature of interactions between the atoms and the underlying confining potential. The dipolar atoms experience repulsion in the x-y plane and attraction along the polarization axis. To minimize repulsion and, consequently, the total energy, it is energetically favorable for the atoms to occupy the ring with a larger radius. However, placing all the atoms there would enhance both the potential energy and the total energy of the system. Consequently, the density in the inner ring remains non-zero. The smaller the value of  $\epsilon_{\rm dd}$, the more particles it contains.

The imbalance of population also determines if (and how) the density modulation emerges.  For smaller $\epsilon_{\rm dd}$, the system is expectedly in the SF regime. A representative 3D density isosurface of such state at $\epsilon_{\rm dd} = 1.95$ is shown in Fig.~\ref{fig_density}(a). (Note that here, the phase is constant). Earlier studies in a simply-connected confinement potential have demonstrated that at a fixed interaction strength, for increasing particle number  the formation of the SS state is favored~\cite{Boettcher2019,Tanzi2019a,Chomaz2019}.  Owing to its larger population, the condensate in the outer ring becomes more prone to a periodic density modulation  (similar to the one in Fig.~\ref{fig_density}(c), but with uniform phase) as $\epsilon_{\rm dd}$ increases. 
For a quantitative analysis of the phase transition, we calculate the superfluid fraction, $f_s=1-I/I_c$, where $I=\text{lim}_{\Omega\rightarrow 0}L/\Omega$ \cite{Leggett1970, Leggett1998}. The classical moment of inertia $I_c=M\langle r^2\rangle$ 
is obtained from the ground state. The $f_s$ as a function of $\epsilon_{\rm dd}$ is shown in Fig.~$\ref{fig_n2_fs}(a)$ for the parameters specified above. For $\epsilon_{\rm dd} \lesssim   2.0$, we have $f_s =1$, as expected for the SF. For larger values of $\epsilon_{\rm dd}$, the ground state is a SS, with the outer ring showing nine density maxima for dipolar strength in the interval  $2.15\geq\epsilon_{\rm dd}\gtrsim 2.078$. The number of localized density sites can be systematically controlled by scaling both \( N \) and \( r_0 \) such that \( N / r_0 \) remains constant, while all other parameters are kept fixed. We note that considering larger $r_{0}$, and decreasing $r_2 - r_1$, the density modulation may occur in both rings. However, we restrict our analysis to the scenario of density modulation forming only in the outer ring.
We also note that for  atoms with only short-range interactions it is possible to have almost equal number of atoms in both rings~\cite{Borysenko2024}, but intrinsic density modulations forming a SS do not develop in these systems.

Examining the role of the Gaussian barrier, we find that, within the range $V_{ B} =(10^4-10^5)a^2_{\rm dd}$, its impact on the superfluid fraction and the corresponding phase transition is weak.
 Specifically, 
 for $V_{B}=10^4a^2_{\rm dd}$, the eight-fold modulated state becomes energetically favorable for $2.15\geq\epsilon_{\rm dd}\gtrsim 2.078$. A smaller value of $V_{B}$ leads to a density overlap between the rings along the radial direction; see the inset of Fig.~\ref{fig_n2_fs} where we have shown the density at the position of barrier for varying barrier strength $V_{B}$.  As we will discuss in the subsequent section, when the system is set into rotation, such finite density at the azimuthal barrier significantly influences the vortex position and thus the angular momentum of system.  \\

\subsection{Rotational ground states:}\label{rot_states}
Let us now investigate the rotational properties of the system to understand how the inter-ring connection determines which part of the system acquires angular momentum and how it influences the nucleation of topological defects.
\subsubsection{Separated rings }\label{sep_rings}
\begin{figure}[h!]
\begin{center}
\includegraphics[width=0.48\textwidth]{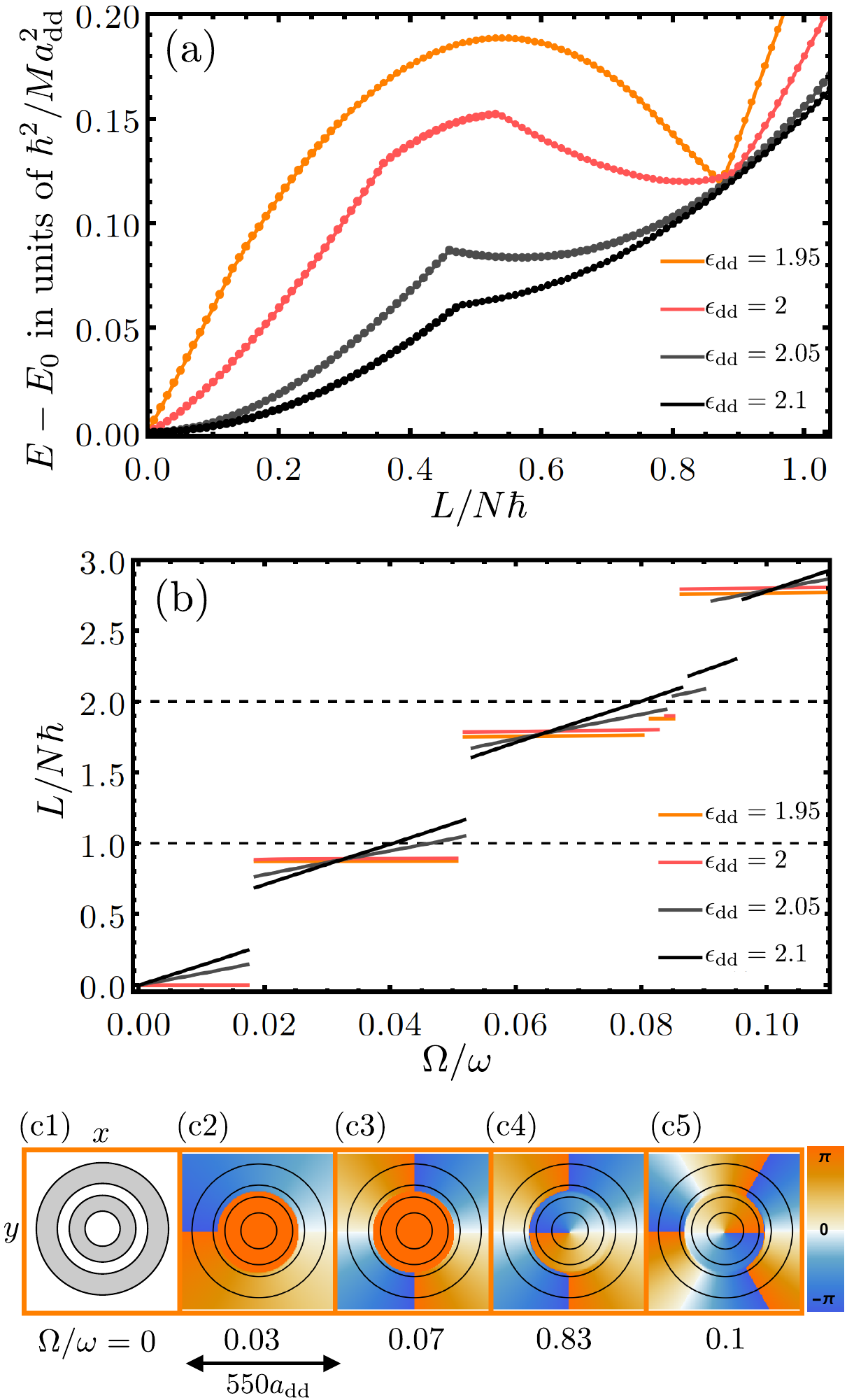}
\end{center}
 \caption{Rotational properties of the double-ring dipolar system with the barrier strength $V_{ B} = 10^5 a^2_{\rm dd}$ for different values of interaction parameters $\epsilon_{\rm dd}$ [see the legends].
 (a) Ground state $E(L)$ relative to the non-rotating ground state $E_0$ as a function of total angular momentum $L$ of the system. (b) Angular momenta $L$ as a function of rotation frequency $\Omega$. The dashed lines indicate the integer values of $L$ expected for a central vortex. (c1)-(c5) The two-dimensional phase profile in the plane of the rings [at $z =0$] for different values of  $\Omega$ for a specific $\epsilon_{\rm dd}=1.95$. The black line indicates the density contours taken at a value of $1/30$ of the maximum density corresponding to each profile. The rotation frequencies are given below each plot. For (\(c_1\)), the phase is constant, and the regions where the inner and outer condensates are present are shaded in grey for clarity. The corresponding states are: (c1) no vortex; (c2) JV$_1$; (c3) 2 JV$_1$s;(c4) CV + JV$_1$; and (c5) 2JV$_1$s + CV}.
\label{rot_fig_sep}
\end{figure}

We first discuss the case in which the rings are completely separated by a barrier of strength $V_{ B} = 10^{5}a^2_{\rm dd}$.
By analyzing the energy as a function of angular momentum [Fig.~\ref{rot_fig_sep}(a)], the angular momentum as a function of rotation frequency [Fig.~\ref{rot_fig_sep}(b)], and the spatial phase distributions [Figs.~\ref{rot_fig_sep}(c1)--(c5)], we can effectively characterize the existence of persistent currents in the ground state,  and the vortex nucleation. 
For the SF state at $\epsilon_{\rm dd} = 1.95$, the yrast line $E(L)$ exhibits the characteristic downward cusp, i.e., a V-shaped minimum, at $ L/N \hbar \approx 0.87 $, identifying the state that can host a persistent current [orange line in Fig.~\ref{rot_fig_sep}(a)]. This is further validated by calculating the ground state in the rotating frame, and analyzing the $L$ vs. $\Omega$ behavior, which exhibits a sudden jump in angular momentum to $L \approx 0.87 N \hbar$ at the critical rotation frequency $\Omega = 0.018\omega$, as shown in [Fig.~\ref{rot_fig_sep}(b)]. Notably, this value matches with the fraction of particles in the outer ring $N_2/N = 0.87$, indicating that angular momentum is  carried by it. This becomes evident from Fig.~\ref{rot_fig_sep}(c2), where a representative two-dimensional phase profile at $\Omega = 0.03 \omega$, reveals a uniform phase in the inner ring, while the outer ring exhibits a full $2 \pi$ winding. This suggests that a vortex resides at the junction between the rings. It is indetified as a \(\rm JV_1\) in the rotating ground state.
In the SF state, the wave function of a $\rm JV_1$ located at position \(\mathbf{r}_{\rm JV_1}\) can be expressed as  

\begin{align} \label{JV_WF}
\psi_{\rm JV_1} (\mathbf{r}) = \sqrt{n(\mathbf{r})} \Bigg\{  
\begin{array}{ll}  
e^{i(\phi +\pi)}, & r > r_0 \\  
e^{i \angle(\mathbf{r}_1, \mathbf{r}_{\rm JV_1})}, & r < r_0 \\  
\end{array}\,\, .   
\end{align}  

Here, the term \(\angle(\mathbf{r}_1, \mathbf{r}_{\rm JV_1})\) represents the angle between \(\mathbf{r}_{\rm JV_1}\) and the reference vector \(\mathbf{r}_1 = (1,0)^T\). This expression is only valid when the density at $r=r_0$ vanishes to ensure the continuity of the wave function. The phase of the inner ring is inherently linked to the position of the $\rm JV_1$. Specifically, at the location of the $\rm JV_1$ the phase of the wave function has a jump of $\pi$ in the radial direction.

Energy minima  in the yrast line [only one is shown here for brevity] correspond to a metastable state where persistent current can be created within a specific ranges of rotation frequencies. For instance, the next  state, appearing for $\Omega>0.051\omega$, has $L =2N_2\hbar$, and accommodates two $\rm JV_1$s [Fig.~\ref{rot_fig_sep}(c3)]. A particularly intriguing transition occurs at $\Omega = 0.081\omega$, when one $\rm JV_1$ migrates to the center, giving rise to a state that hosts both a $\rm JV_1$ and a CV [Fig.~\ref{rot_fig_sep}(c4)]. In this configuration, each particle in the inner ring acquires superfluid circulation, leading to $L = (N_2+N)\hbar$. Similarly, by further increasing the $\Omega$, it is possible to create a system that supports multiples $\rm JV_1s$ and CVs [see Fig.~\ref{rot_fig_sep}(c5) for two $\rm JV_1$s and one CV]. 

In the SS state (\(\epsilon_{\rm dd} > 2.078\)), the minimum of the yrast line shifts to a lower value of \(L/N\hbar\). For smaller angular momentum the variation of energy is more parabolic in nature for larger \(\epsilon_{\rm dd} \) in contrast to the linear variation that we observe in SF. Moreover, a kink appears at \(L = N_2 \hbar/2\) due to the intersection of two energy branches arising from the system’s kinetic energy. Notably, while in a single-ring SS, the kink location at \( L = N\hbar/2 \) remains unchanged and is determined solely by the total number of particles, in a double-ring system it shifts towards \( N/2 \) as the outer ring progressively becomes more populated for larger $\epsilon_{\rm dd}$; see  \(\epsilon_{\rm dd} = 2.05\) and \(2.1\) in Fig.~\ref{rot_fig_sep}(a). The energy barrier that prevents the metastable state from decaying into the non-rotating state also depends on $f_s$; compare the range between $\epsilon_{\rm dd} = 2.1$ and $2.05$. Additionally, in the angular momentum of the ground states, we observe a gradual slope linked to the value of \( f_s \) before it abruptly jumps to a higher value with increasing rotation frequency.  While the rotational states for different \(\epsilon_{\rm dd}\) exhibit the same number of $\rm JV_1s$ and CVs, their critical rotation frequencies differ  due to a varying population in the outer ring. If the superfluid fraction drops below a critical threshold, vortices can still nucleate in the rotating ground state; however, they do not generate a persistent current, causing the system to decay into a non-rotating state in the dynamics once $\Omega$ is reduced to zero.

Near the phase transition in the non-rotating ground state on the SF side, we observe that rotation can induce density modulation, driving the system into the SS phase. This effect is evident for \(\epsilon_{\rm dd} = 2\). The yrast line reveals that the condensate remains in the SS phase within the interval \( 0 < L/N\hbar < 0.35\) and for \(L/N\hbar>0.54\). However, when the energy connected to the rigid body rotation of the SS becomes large,
it is energetically favorable to transform the system back to a SF state. Thus, the yrast line retains a concave segment for \( 0.36 < L/N\hbar < 0.54 \). But the metastable state at \( L/N\hbar = 0.82 \) lies in a SS phase hosting a $\rm JV_1$. The density modulations, though present in the outer ring, are not pronounced  enough (compared to others) to generate a discernible slope in the \( L \) vs. \( \Omega \) plot in Fig.~\ref{rot_fig_sep}(b). 

\begin{figure}
\begin{center}
\includegraphics[width=0.48\textwidth]{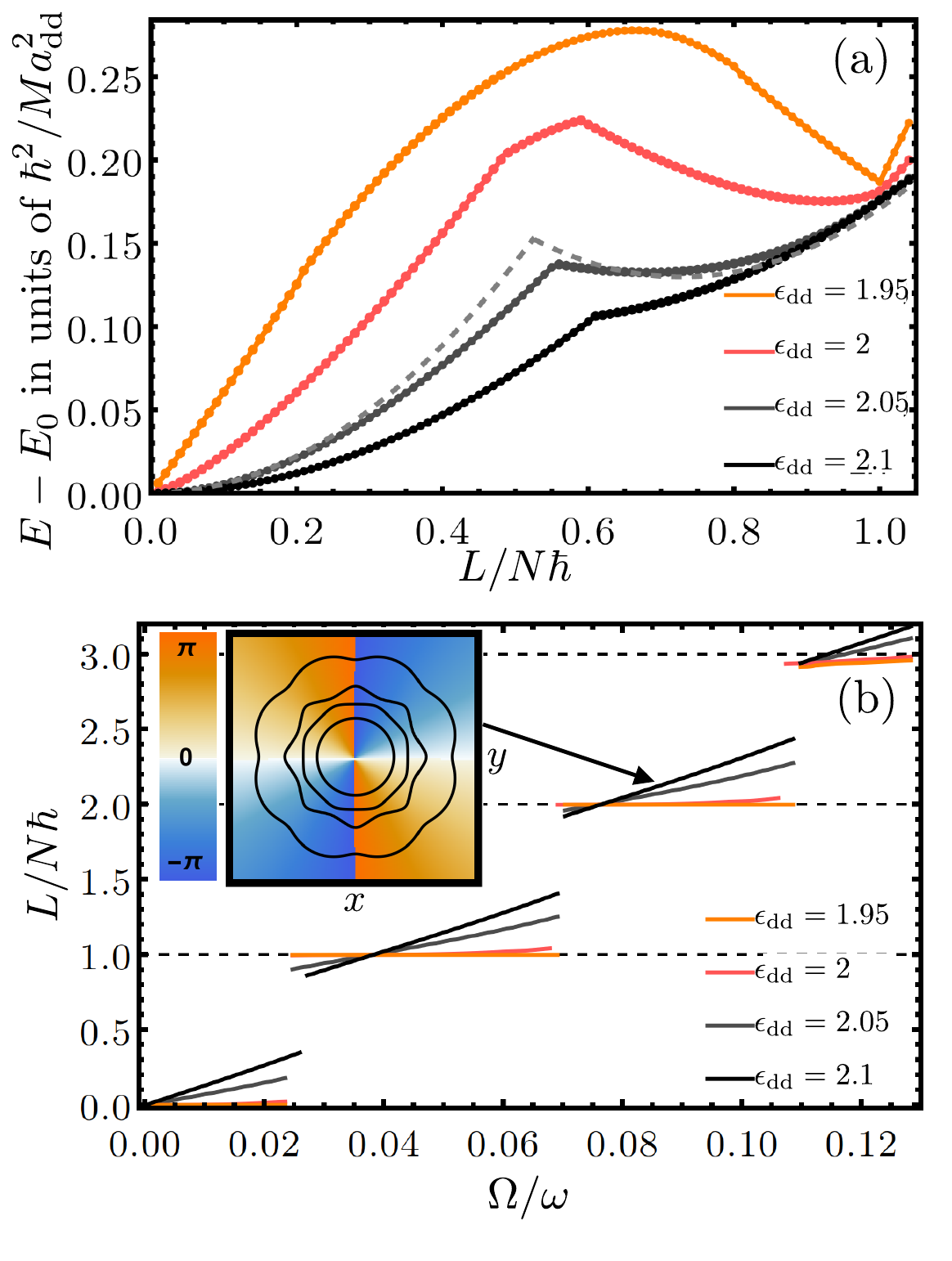}
\end{center}
\caption{Rotational states of double-ring dipolar system with a barrier strength $V_{B} = 10^4 a^2_{\rm dd}$ for different $\epsilon_{\rm dd}$ [see the legends] (a) Energy dispersion relation $E(L)$ relative to the non-rotating ground state $E_{0}$ as a function of angular momentum per particle. The points are calculated from the fixed angular momentum routine and the dashed line shows Eqs. \eqref{Yrastss}, where we have used the state $\psi(L=0)$ for the expectation value.(b) Angular momentum $L$ of the rotational ground state realized at various rotation frequencies $\Omega$. The inset shows the phase profile with the density contour [value being $1/30$ times the maximum density] for $\epsilon_{\rm dd}=2.1$ at a specific rotation frequency $\Omega = 0.09 \omega$ corresponding to 2CVs. }
\label{rot_fig_con}
\end{figure}

\begin{figure*}
\begin{center}
\includegraphics[width=1\textwidth]{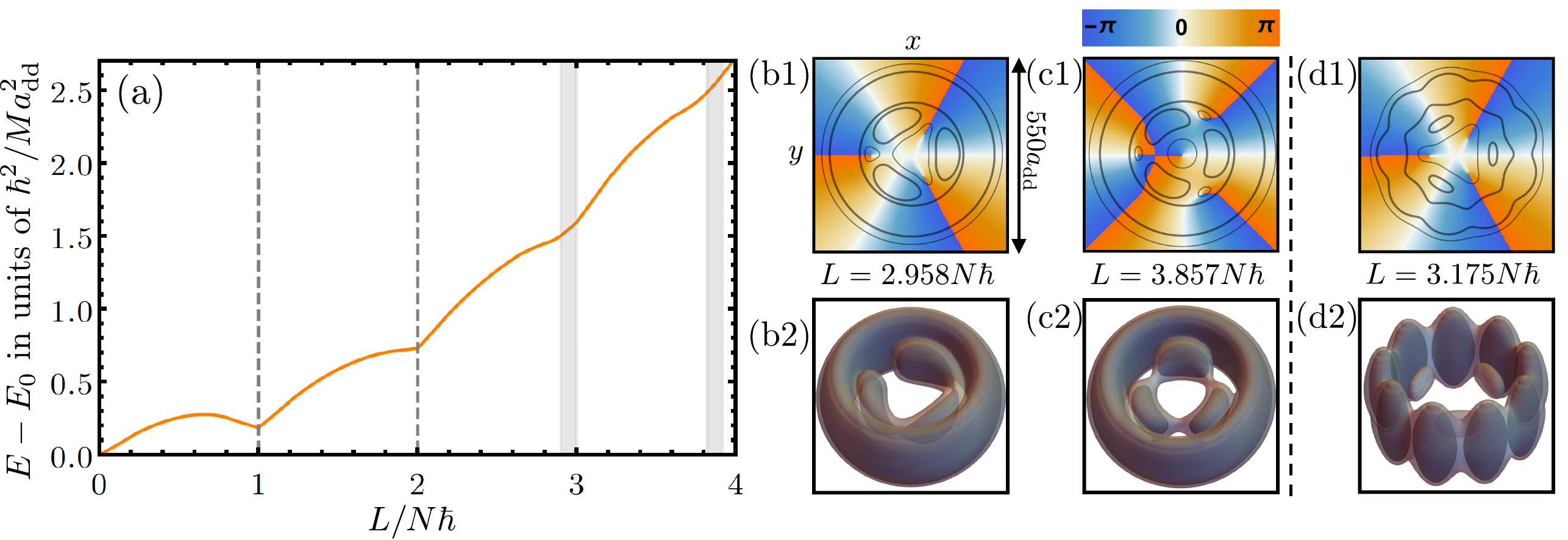}
\end{center}
\caption{The formation of density modulation in the inner ring and the SF-SS phase transition induced by rotation for $V_{B}=10^4a_{\rm dd}^2$ (a) Energy $E(L)$ as a function of angular momenta $L$ relative to the non-rotating ground state energy $E_{0}$ for $\epsilon_{\rm dd}=1.95$.  The dashed lines indicate the angular momenta of the global minimum in the rotating frame, where the ground states are in the SF phase. The gray-shaded area represents all angular momenta corresponding to the SS ground states. Shown also the phase profiles in the plane $z = 0$ [(b1), (c1), (d1)] and three dimensional density isosurfaces [(b2), (c2), (d2)] for $\epsilon_{\rm dd}=1.95$ [(b1), (b2), (c1), (c2)] with (b1)-(b2) $L/N\hbar = 2.958$ (3 JV$_2$s) and (c1)-(c2) $L/N\hbar = 3.857$ (CV + 3JV$_2$s), and (d1)-(d2) for $\epsilon_{\rm dd}=2.1$ with  $L/N\hbar = 3.175$ (3 JV$_2$s). The density contour lines  correspond to a value of $20\%$ and $1\%$ of the maximum density [(b1), (c1), (d1)]. The isosurfaces are taken at $10\%$ and $20\%$ of the maximum densities in (b2) and (c2) and at $3\%$ and $10\%$  of the maximum density in (d2)}.
\label{YrastFull}
\end{figure*}
\subsubsection{Connected rings}\label{con_rings}
The observation of $\rm JV_1$ becomes possible because the density drops to zero at the location of a strong barrier. A weak barrier that maintains non-zero density between the rings favors the formation of a vortex at \( r < r_0 \). The overall behavior of the yrast lines, which support persistent currents at their minima and display kinks in the SS regime, is similar to the case of separated rings [compare Fig.~\ref{rot_fig_sep}(a) and Fig.~\ref{rot_fig_con}(a)]. In particular, we highlight the behavior for $\epsilon_{\rm dd} = 2.1$ when $f_s = 0.490$. The absence of a minimum in the yrast line indicates that persistent currents cannot form in this system. Nevertheless, the rotational ground states can still host vortices [ Fig.~\ref{rot_fig_con}(b)].

Depending on the rotation frequency \(\Omega\), the vortex may manifest either as a CV or as a different type of JV distinct from $\rm JV_1$.  That the vortex is a CV  within the interval \(0.024\omega < \Omega < 0.108\omega\) is evident from the integer value of the angular momentum, shared by all particles in the condensate,  for \(\epsilon_{\rm dd} = 1.95\) [see Fig.~\ref{rot_fig_con}(b)]. In the SS regime,  the same vortex remains a CV, though with a reduced superfluid angular momentum.
The critical rotation frequencies for the system at which the charge of the CV changes from $q-1$ to $q$ are given  as the solutions of the transcendental equation 
\begin{eqnarray}\label{Omega}   
\Omega = \frac{\hbar\langle r^{-2}\rangle}{M}\left( q-\frac{1}{2}\right)\,\, .
\end{eqnarray}
We note that the mean value  is calculated from the wave function which, in general, depends on all parameters of the system. 

Let us now analyze the shape of the yrast line. The $E$ is linear for small value of $L$ and around the minima. We can understand the behavior at small $L$ by assuming that the vortex is far from the condensate, such that its core does not interfere with the condensate density. In this scenario, the wave function can be approximated as

\begin{align}
\label{Vortex_Eqn}
\psi = \sqrt{n(\mathbf{r})}\exp\left[i \tan^{-1} \left( \frac{y}{x - x_0} \right) \right]\,\, ,
\end{align}

where we assume the vortex to enter on the x-axis, such that its position in the xy-plane is given by $(x_0,0)$, and $n(\mathbf{r})$ is the local density. Inserting Eq.~\eqref{Vortex_Eqn} in the energy functional and expanding linearly around $L = 0$, we  obtain
\begin{align}
\label{yrast}
E-E_0\approx E_{\rm kin}= \frac{\hbar N L}{2M\langle r^2\rangle}\,\, ,   
\end{align}
which is valid as long as $x_0^2\gg \langle r^2 \rangle$. 
To understand the linear behavior near the minimum energy, where the vortex is located at the center and the interaction between the vortex and condensate can be neglected,  we again take Eq. \eqref{Vortex_Eqn} as an ansatz and expand linearly around $L = N \hbar$,
\begin{align}
\label{yrast2}
E-E_0\approx E_{\rm kin}= \frac{\hbar^2}{2M}\langle r^{-2}\rangle +\frac{\hbar}{M}(L-N\hbar)\frac{\langle x/r^4\rangle}{\langle x/r^2 \rangle}\,\, ,   
\end{align}
For arbitrary values of $L$, the energy of the vortex consists of two parts: the kinetic energy and the interaction energy which arises from the density depletion when the vortex penetrates the condensate. The latter is a concave parabolic function of $L$. While Eqs.~\eqref{yrast} and ~\eqref{yrast2} can qualitatively capture the linear behavior in the SF state, a quantitative agreement requires considering the deep SF regime or a non-dipolar superfluid characterized by dominant short-range interactions. For a SS the situation is entirely different. An additional term in the kinetic energy emerges from the solid body rotation, reading as $E_{\rm SS}=L^2/2I_{\rm sb}$ where $I_{\rm sb}=(1-f_s)M\langle r^2\rangle$. The contribution of the interaction energy decreases with decreasing $f_s$, because the vortex can pass through interstitial region between the density maxima and therefore minimizing the interaction.

In order to calculate the yrast line for the SS state, let us decompose the total angular momentum $L$ into two parts, namely a SF part $L_{\rm SF}=f_{s}N\hbar$ and a SS part $L_{\rm SS}=L-L_{\rm SF}$. The yrast line can then be constructed by 
\begin{align}
\label{Yrastss}
E(L)=\text{min}\left[E_{\rm SS}(L),E_{\rm SS}(L_{\rm SS})+f_s\frac{\hbar^2}{2M}\langle r^{-2}\rangle\right]\,\, .
\end{align}
In Figs.~\ref{rot_fig_con}(a), the E(L) is plotted (see the dashed line) using Eq.~\eqref{Yrastss} for $\epsilon_{\rm dd}=2.05$ , where we have calculated the corresponding expectation values using $L = 0$. The function [Eq.~\eqref{Yrastss}] agrees well with the exact yrast line. This shows that the yrast line of a supersolid is determined by the superfluid fraction together with the particle density, and the interactions play a minor role. The yrast line has a kink at 
\[L/N\hbar = f_s/2 +(1-f_s)\langle r^2\rangle \langle r^{-2}\rangle/2N^2\,\, .\] Further, the yrast line has a metastable state at $L=f_sN\hbar$ as long as $f_s> f^{c}_s$, where $f^{c}_s = \langle r^2\rangle\langle r^{-2}\rangle/(N^2+\langle r^2\rangle\langle r^{-2}\rangle)$. For the $\epsilon_{\rm dd} = 2.1$, $f^c_s = 0.533$ and $f_s = 0.493$, and therefore no-metasble state is observed in Fig.~\ref{rot_fig_con}(b). We also point out that all equations mentioned above are also valid for the yrast lines of the separated ring cases if we exchange $N$, $\langle ... \rangle$ and $f_s$ by $N_2$, $\langle ... \rangle_2$ and $f_{s,2}$, where $\langle ... \rangle_2$ means integration over the outer ring only and $f_{s,2}$ is the superfluid fraction of the outer ring. The latter can be obtained by calculating the angular momentum and the moment of inertia for the outer ring only. Furthermore, for a single-ring system with tight confinement and localized density, the relation 
\(\langle r^2 \rangle \langle r^{-2} \rangle \sim 1\) 
holds, indicating that the kink appears at \( L/N\hbar = 1/2 \). In contrast, for the double-ring system, the kink position can be varied significantly by tuning \(\epsilon_{\rm dd}\) and \( V_{B} \), thereby allowing the system to host a wider range of vortex configurations. \par

The most intriguing effects of the connected-ring geometry emerge at high angular momentum states. At these higher angular momentum states, rotation alone induces density modulation in the inner condensate, transforming it into a SS state with three localized density sites. The presence of SS is reflected in the slope of the angular momentum, detectable even for $\epsilon_{\rm dd} < 2$; see Fig.~\ref{rot_fig_con}(b).  To gain deeper insight into this phenomenon, we calculate the yrast line up to high angular momentum states, $L/N\hbar = 4$, for $\epsilon_{\rm dd} = 1.95$ [Fig.~\ref{YrastFull}(a)]. Two types of topological defects can be identified in the dispersion relation. The first two dashed lines, indicating kinks at $L/N\hbar = 1,\,2$, correspond to CVs. In any rotating frame, $E - L\Omega$, the positions of these kinks remain fixed.  Additionally, we observe two defects with angular momenta indicated by the gray regions. In a rotating frame, these defects can form a global minimum at values $L_0$, which satisfy $\left. \partial E / \partial L \right|_{L=L_0} = \Omega$, meaning that the position of the minimum depends on the rotation frequency at which the vortices are nucleated. This behavior is possible only if solid-body rotation occurs due to the formation of localized density sites, with the associated vortices being $\rm JV_2s$, which form at the junctions between these sites.  
The ensuing sites and three $\rm JV_{2}s$ at the junctions between these sites for $\epsilon_{\rm dd} = 1.95$ and $\epsilon_{\rm dd} = 2.1$ are shown  in Figs.~\ref{YrastFull}(b1)-(b2) and Figs.~\ref{YrastFull}(d1)-(d2), respectively. These are the stationary states for fixed value of the angular momenta. The one with $\epsilon_{\rm dd} = 2.1$ correspond to the SS with a density modulation in both inner and outer rings. We show that adding the next vortex as a CV is energetically favorable  if the angular momentum becomes larger such as $L/N \hbar = 3.857 $ [Figs.~\ref{YrastFull}(c1)-(c2)], and thus, enabling the coexistence of CV  and  $\rm JV_2$.
\label{phase_dia}
\begin{figure*}[t!]
\begin{center}
\includegraphics[width=1\textwidth]{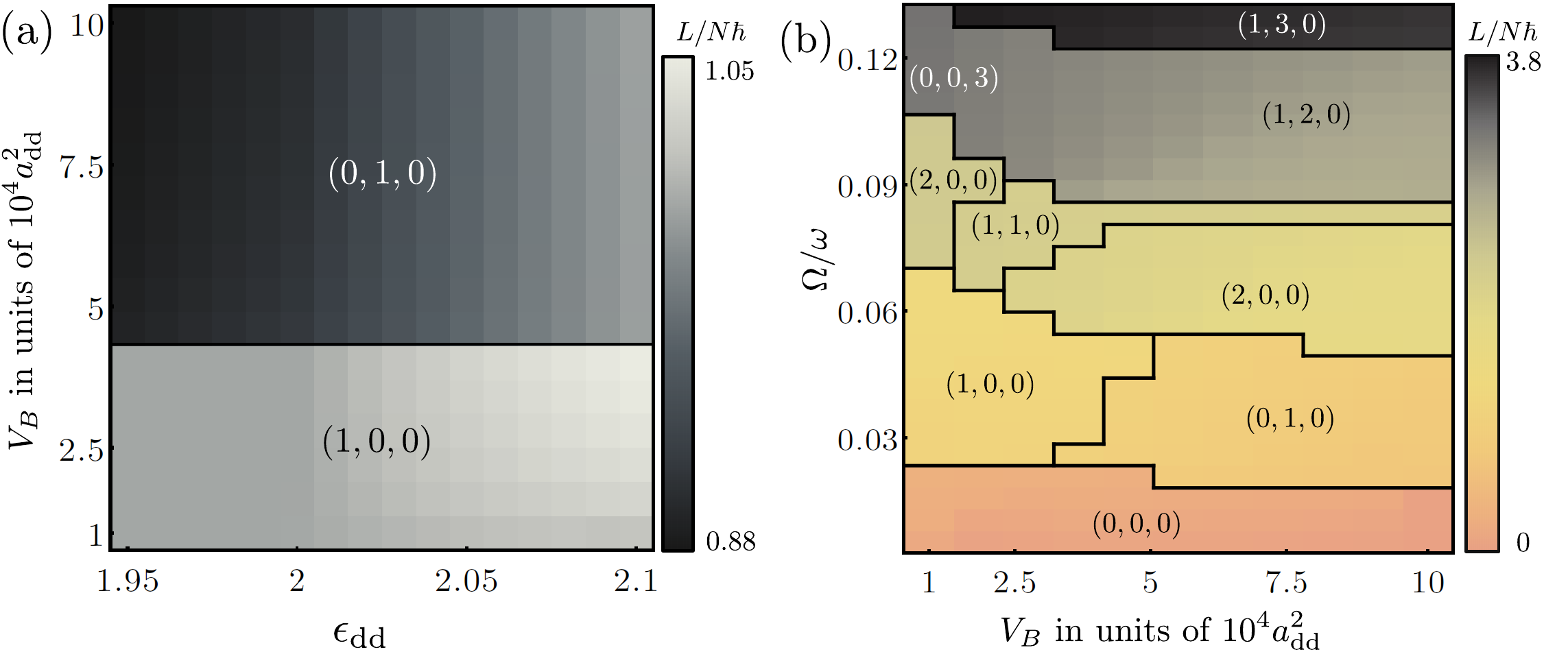}
\end{center}
\caption{
Angular momentum (a) as a function of barrier strength $V_{B}$ and interaction strength $\epsilon_{\rm dd}$ at a fixed rotation frequency $\Omega = 0.04\omega$, and (b) as a function of barrier strength $V_{B}$ and rotation frequency $\Omega$ at fixed $\epsilon_{\rm dd} = 2$. Each state in the phase diagram is labeled by the charges $q_i$ of the topological defects $(q_{\rm CV}, q_{\rm JV_1}, q_{\rm JV_2})$.}

\label{PhaseDia}
\end{figure*}
We note that the rotation-induced SS also occurs in a single-ring configuration~\cite{Tengstrand2021} and arises from the fact that, in the rotating frame, the energy at the roton minimum can satisfy the condition \( E_{\rm rot} - \Omega L \leq 0 \) by tuning \(\Omega\). Notably, this provides an alternative protocol for generating an SS state, independent of controlling \(\epsilon_{\rm dd}\). This mechanism is particularly relevant in our double-ring system, where modulation can appear in either the outer or inner ring, depending on the presence of inter-ring density connection. The emergence of modulation in the inner ring is intrinsically linked to the $\rm JV_2s$ and represents one of the key highlights of our work.
\subsubsection{Phase-diagrams}
We present the nucleation of distinct topological structures as a function of system parameters, $V_B$, $\epsilon_{\rm dd}$, and $\Omega$ delineated in the two phase diagrams shown in Figs.~\ref{PhaseDia}. The parameters, $N$, $(\omega, \omega_z)$, and $r_{0}$ are kept fixed. The total charge associated with each vortex type is denoted by $q_c$, where $c \in \{\mathrm{CV}, \mathrm{JV1}, \mathrm{JV2}\}$. A given configuration is represented by the charge tuple $(q_{\text{CV}}, q_{\text{JV1}}, q_{\text{JV2}})$.
\label{vor_det}
\begin{figure*}[t!]
\begin{center}
\includegraphics[width=0.99\textwidth]{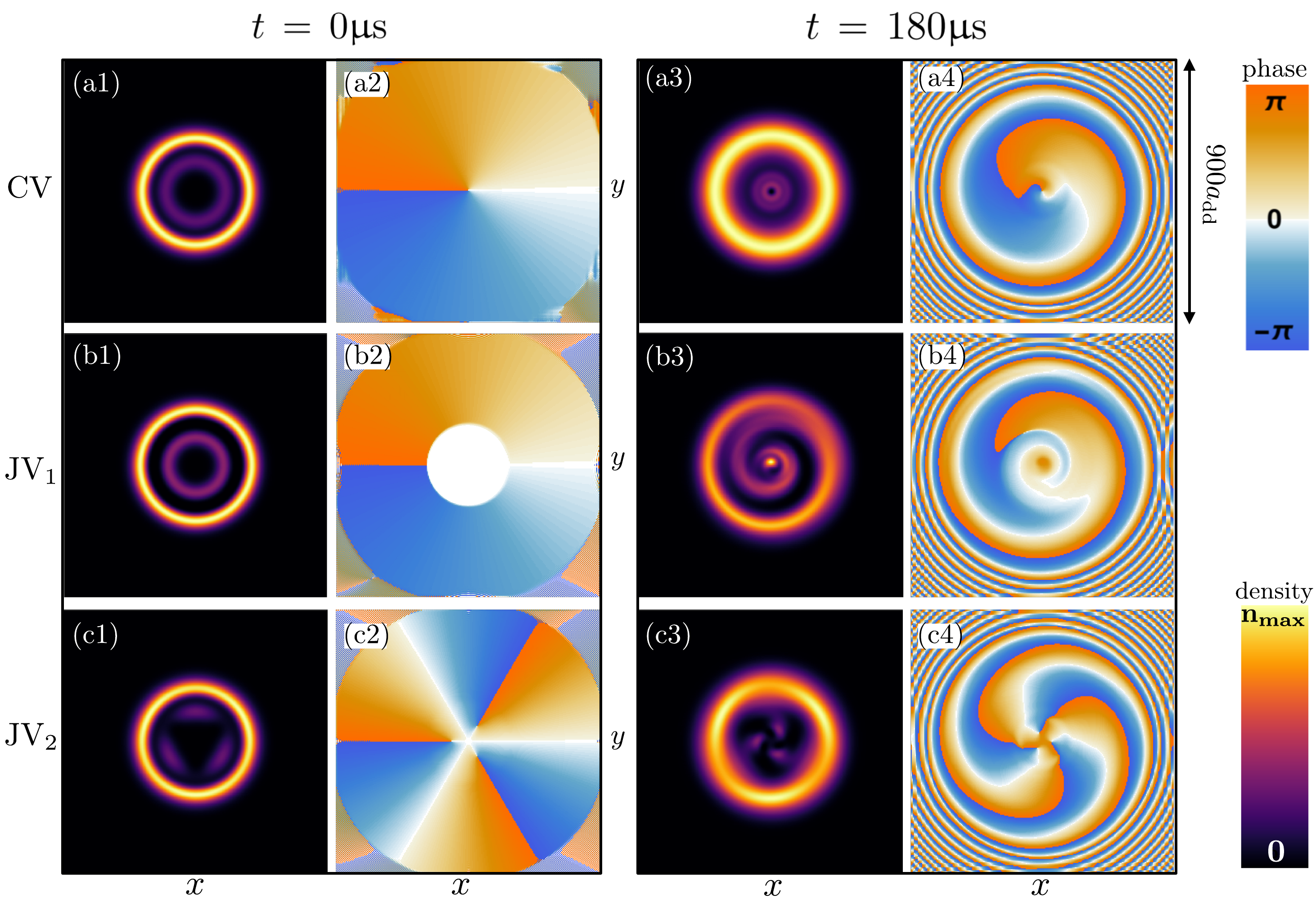}
\end{center}
\caption{Interferometric protocol for detection of (a1)-(a4) CV, (b1)-(b4) $\rm JV_1$, and (c1)-(c4) $\rm JV_2$. Shown here two dimensional density $n_{\rm 2D}(x, y)$ [(a1), (b1), (c1), (a3), (b3), (c3)] and phase profiles [(a2), (b2), (c2), (a4), (b4), (c4)] at $z = 0$ in the dipolar double-ring system at $t = 0$ [(a1)-(a2), (b1)-(b2), (c1)-(c2)] and $t = 180 \mu s$ [(a3)-(a4), (b3)-(b4), (c3)-(c4)]. After preparing the initial states all the confinement potentials are set zero for $t>0$. The initial condensates have an interaction strength of $\epsilon_{\rm dd} =1.95$, rotation frequencies of $\Omega/\omega = 0.04,\,0.04,\, 0.1$ and $V_ B=10^4a_{\rm dd}^2,\,10^5a_{\rm dd}^2,10^5a_{\rm dd}^2$ for CV, JV$_1$ and JV$_2$.
}
\label{fig_dyn_intf}
\end{figure*}
First, we examine how the Gaussian barrier influences the presence of $\rm JV_1s$ and CVs for different interaction strengths $\epsilon_{\rm dd}$. We fix the rotation frequency at $\Omega = 0.04\omega$ and calculate the total angular momentum as a function of $V_{B}$ and $\epsilon_{\rm dd}$, as shown in Fig.~\ref{PhaseDia}(a).  For the chosen value of $\Omega$, CVs exist when $L \geq N\hbar$, while $\rm JV_1s$ are present when $L < N\hbar$. We observe a critical barrier strength $V_{B,c} \approx 3.3 \times 10^4 a_{\rm dd}^2$ that separates the two regions, $(0,1,0)$ and $(1, 0, 0)$, of the phase diagram indicating a single $\rm JV_1$ and a single CV. By comparing this with the inset in [Figs.~\ref{fig_n2_fs}] we see that the density for $V_{B,c}$ is non-zero. Consequently, the creation of the $\rm JV_1$-core requires kinetic energy, but still the total energy of the $\rm JV_1$ is smaller than the total kinetic energy of a CV. For the parameter grids in [Figs.~\ref{PhaseDia}(a)] we do not see a dependency of $\epsilon_{\rm dd}$ on $V_{ B,c}$. The SF-SS phase transition is also evident in Fig.~\ref{PhaseDia}. For $\epsilon_{\rm dd} \leq 2$, in the SF state, the angular momentum of the CV is constant and equal to $N\hbar$. In the SS phase, the angular momentum of the CV is no longer constant due to the contribution from the solid part of the system, which increases with larger $\epsilon_{\rm dd}$.  In the case of a $\rm JV_1$, this effect is reinforced by the increasing population of the outer ring. This result once again confirms that the barrier is the cause for the nucleation of $\rm JV_1s$.\\
\indent Next, we fix the interaction strength to $\epsilon_{\rm dd} = 2$ and calculate the angular momentum of the system for different values of $\Omega$ and $V_B$ to identify various vortex configurations; see Fig.~\ref{PhaseDia}(b). Typically,  CVs are energetically favored at low barrier strengths \( V_B \), while JV\(_1\) states become preferred as \( V_B \) increases, see, for example, in the regime (\( 0.03 < \Omega/\omega < 0.05 \)). This behavior is consistent with the Figs.~\ref{rot_fig_sep}, \ref{rot_fig_con}, and \ref{PhaseDia}(a). As the rotation frequency increases, a JV\(_1\) can coexist with a CV even at relatively small \( V_B \). Notably, the minimal barrier strength required for such a configuration, \((1,1,0)\), is \( V_{B,c} \approx 1.9 \times 10^4 a^2_{\rm dd} \) [Fig~.\ref{PhaseDia}(b)]. We notice that transitions between configurations with different numbers of JV\(_1\) and CV excitations occur sharply as a function of both \( V_B \) and \( \Omega \), due to the clear distinguishability of their preferred locations in the double-ring system.  Remarkably, the above-mentioned transition is no longer sharp when the system involves a \( \rm JV_2 \); see, for example, the angular momenta as the system crosses over from the \( (0,0,3) \) to the \( (1,2,0) \) state. In this regime, as the barrier strength is gradually increased for a fixed \( \Omega \), say \( 0.12\, \omega \), two of the three \( \rm JV_2 \) excitations move to \( r = r_0 \), while one \( \rm JV_2 \) moves to \( r = 0 \). For \( \epsilon_{\rm dd} = 2 \), the non-rotating ground state lies close to the phase transition on the SF side. While nucleating the \( (0,0,3) \) configuration, the inner ring develops a density modulation, which steadily fades as \( V_B \) increases. Consequently, the total angular momentum gradually decreases from approximately \( 2.96\,N\hbar \) to \( 2.81\,N\hbar \). Eventually, in the strong barrier regime, this vortex configuration evolves into the \( (1,2,0) \) state.

\subsection{Vortex Detection}
In the preceding sections, we have demonstrated how JVs and CVs can be nucleated in a double-ring system. This system is particularly advantageous as it provides a practical method for detecting these vortices once they are generated. The corresponding protocol involves interfering condensate parts located in the inner and outer rings by switching off the trap and monitoring the resulting interference pattern. Since JVs and CVs exhibit distinct phase profiles, they are expected to produce discernible density patterns during expansion. We illustrate such expansion dynamics in Fig.~\ref{fig_dyn_intf} focusing only SF with $\epsilon_{\rm dd} =1.95$ . 

The initial 2D  density profiles $n_{\rm 2D}(x, y)$ and phase profile for a CV are highlighted in Fig.~\ref{fig_dyn_intf}(a1)-(a2), respectively. Such initial state is created with $V_{B} =10^4a^2_{\rm dd}$ and $\Omega = 0.04 \omega$. Here, both the inner and outer condensates contain one complete phase winding, resulting in concentric circular density patterns after interference [Figs.~\ref{fig_dyn_intf}(a3)-(a4)]. Such density and phase profiles have to be contrasted when a $\rm JV_1$ is present in the system for $V_{B} =10^5a^2_{\rm dd}$;
see Figs.~\ref{fig_dyn_intf}(b1)-(b4). The inner ring posses a uniform phase whereas the phase winds over $2 \pi $ in the outer ring owing to single $\rm JV_1$. Consequently, after interference, the density develops a single spiral around the center. However, the presence of density modulation in the inner ring, which is otherwise an SF in the absence of rotation, indicates the existence of \( \rm JV_2s \). This is further confirmed by the interference pattern, where we observe that modulated density structures persist in the inner region of the condensate, with the \( \rm JV_2s \) located between these structures [Figs.~\ref{fig_dyn_intf}(c3)-(c4)]. We remark that, while our results on interference dynamics are illustrated using a specific set of parameters and focus on short timescales to avoid boundary effects, the phenomena remain observable over longer durations and are generally applicable, with the timescales being adjustable via the trapping frequencies.Thus, this interference protocol not only confirms the existence of a defect but also reveals the phase distribution resulting from its specific position.

\section{Conclusions}\label{conclu}
In conclusion, we have investigated the non-rotational and rotational properties of dipolar atoms in coplanar and concentric double rings. We have specifically focused on two different cases: one where the rings are connected by density overlap and another where they are separated. Our findings reveal a population imbalance between the inner and outer rings, in contrast to non-dipolar systems. Notably, the outer ring exhibits a proclivity for spontaneous density modulation, indicative of a SS state, regardless of the strength of the barrier forming the double-ring structure.

We have studied the rotational properties of the system and demonstrated the existence of a persistent current by calculating its energy as a function of angular momentum. The persistent current is accompanied by the formation of topological structures, specifically $\rm JV_1$ and $\rm CV$, which appear in the separated and connected rings, respectively. We have delineated their regions of existence in a diagram by calculating angular momentum as a function of interaction and barrier strength. Specifically, we have identified multiple topological defects, such as $\rm JV_1s$ and combination of $\rm CV$ and $\rm JV_1$, which emerge in the separated ring case at higher rotational frequencies.

One of the intriguing features of our system lies in its connection to rotation-induced SS state formation. We have found that, for relatively small angular momentum, rotation can induce density modulation in the outer ring, transforming an otherwise SF state into an SS state. Interestingly, in the case of connected rings, the inner condensate forms three localized density sites at high angular momentum states, where three vortices are nucleated at the junctions between the localized density sites. These vortices, which we refer to as $\rm JV_2$, are associated with the spontaneously formed localized density sites induced by rotation and are unique to our double-ring dipolar system. Finally, by utilizing an interferometric protocols, we have demonstrated how these different topological structures can be detected in the experiment. 

There are several extensions of the present work worth pursuing in future research. A straightforward extension would be to study the underlying collective excitation spectra of the double-ring system, particularly around the critical point of the phase transition. Additionally, it would be interesting to explore the signature of $\rm JV_2$ in the excitation spectrum. Another intriguing direction would be to investigate the parameter regime where density modulation occurs in both rings in the non-rotating supersolid state. This would enable the study of shear-wave propagation across the azimuth of the rings by suddenly altering the distance between the inter-ring localized density sites. Furthermore, investigating finite-temperature properties~\cite{Sanchez2022} and the effects arising when dipole moments are tilted with respect to the perpendicular axis in the context of this setup would be equally compelling.

\section*{ACKNOWLEDGMENTS}
We thank Thomas Bland,  Philipp Stürmer, Tiziano Arnone Cardinale, Deepak Gaur, Lila Chergui and Stuttgart  Dipolar Gases group for fruitful discussions.  We also thank  M. Nillson Tengstrand for his help with the eGPE code. M.S., K.M. and S.R. acknowledge financial support from the Knut and Alice Wallenberg Foundation (Grant No. KAW 2018.0217 and KAW2023.0322) and the Swedish Research Council (Grant No. 2022-03654). T.P. acknowledges support from the European Research Council (ERC) (grant agreement No. 101019739).

{\vspace{0.25cm}\noindent\textit{Data Availability}}
The data associated with this work are available from the corresponding author upon reasonable request.

\appendix

\twocolumngrid

\section{Computational Details}\label{comdetails}

Here, we detail the numerical simulations used to obtain the results described in the main text. We numerically solve the 
the extended Gross-Pitaevskii equation (eGPE) obtained from the functional derivative of the energy density function. 
 The eGPE equation is cast into a dimensionless form in our simulations by rescaling the length and the time  by length scale and time scale, $l_s = a_{\rm dd}$ and $t_s = Ma_{\rm dd}^2/\hbar$, respectively. Thereafter, we employ the split-time Fourier spectral method to solve the resulting equation~\cite{BAO2003318, Bao_2003}. The stationary state of the system is obtained through imaginary time propagation, while the dynamical simulation is performed in real time. At each imaginary time step, we preserve the normalization of the wave function, and convergence is reached when the relative deviation of the wave function $\psi(x,t)$ at every grid point and the angular momentum $L$ and  energy $E$ between consecutive time steps are smaller than $10^{-6}$ and $10^{-15}$, respectively. It should be noted that calculating the stationary state solution of the eGPE is an involved task due to many close-lying local minima in the energy surface, which necessitates extensive sampling over many different initial conditions to identify the most probable lowest-energy solutions. The mean-field dipolar potential is efficiently evaluated via Fourier transforms incorporating a spherical cutoff, set to half the box size to prevent spurious interactions between periodic images.
Our simulations are carried out in a 3D box characterized by a grid $(n_x \times n_y \times n_z)$ corresponding to $(128 \times 128 \times 64)$ ($512 \times 512 \times 128$ for the simulation of interferometric protocol). 
The employed spatial discretizations (grid spacing) refers to $\Delta_i = l/n_i $ with $l=900$ for the calculations of rotating and non-rotating ground states ($l=1500$ for the simulations of interferometric protocol), while the time step of the numerical integration is $\Delta_t = 10^{-2}$.

\section{Derivation of Eqs~\eqref{yrast} and \eqref{yrast2}}\label{Derivation}
To arrive at Eqs.~\eqref{yrast} and \eqref{yrast2}, we calculate the angular momentum as a function of  position of the vortex \( x_0 \):
\begin{align}
    L = \langle\hat{L}_z\rangle = \hbar \int dV\, n\, \frac{x^2 - x x_0 + y^2}{x^2 - 2x x_0 + x_0^2 + y^2}\, .
\end{align}
When \( L \rightarrow 0 \), the vortex is far away from the condensate. In this case, the angular momentum can be approximated as \( L = \hbar \langle r^2 \rangle / x_0^2 \). To obtain this expression, we have assumed \( \langle x \rangle = 0 \), which holds since the vortex and condensate are well separated. The kinetic energy can be calculated via Eq.~\eqref{GPEEnergy}. For large \( x_0 \), we obtain \( E(L) - E_0 \approx E_{\rm kin} - E_0 = N \hbar^2 / (2M x_0^2) \), where \( E_0 = E[\sqrt{n}] \). Combining these two expressions yields Eq.~\eqref{yrast}. For \( L \rightarrow 1 \), we expand the angular momentum as \( L = N \hbar + \langle x \rangle x_0 / r^2 \). The expectation value \( \langle x \rangle \) is nonzero because the condensate and vortex are not well separated for angular momenta close to, but not equal to, unity. This holds only if the angular momentum is interpreted as the superfluid angular momentum \( L_{\rm SF} \). Combining this with the kinetic energy expression \( E_{\rm kin} = \hbar^2 \langle r^{-2} \rangle / (2M) + \hbar^2 x_0 \langle x / r^4 \rangle / M \) yields Eq.~\eqref{yrast2}.

\bibliography{reference.bib}

\end{document}